\documentclass{article}

\usepackage{microtype}
\usepackage{graphicx}
\usepackage{booktabs} 

\usepackage{hyperref}

\usepackage[accepted]{icml2021}
\icmltitlerunning{Blind Signal Separation for Fast Ultrasound Computed Tomography}

\usepackage{booktabs}
\usepackage{adjustbox}
\usepackage{subfig}
\usepackage{amsmath}
\begin{document}

\twocolumn[
\icmltitle{Blind Signal Separation for Fast Ultrasound Computed Tomography}



\icmlsetsymbol{equal}{*}

\begin{icmlauthorlist}
\icmlauthor{Takumi Noda}{lm,ut}
\icmlauthor{Yuu Jinnai}{lm}
\icmlauthor{Naoki Tomii}{ut}
\icmlauthor{Takashi Azuma}{lm}

\end{icmlauthorlist}

\icmlaffiliation{lm}{Lily MedTech, Tokyo, Japan}
\icmlaffiliation{ut}{University of Tokyo, Tokyo, Japan}

\icmlcorrespondingauthor{Yuu Jinnai}{ddyuudd@gmail.com}

\icmlkeywords{Deep Learning}

\vskip 0.3in
]


\footnotetext[1]{Lily MedTech, Tokyo, Japan}
\footnotetext[2]{University of Tokyo, Tokyo, Japan}



\begin{abstract}









Breast cancer is the most prevalent cancer with a high mortality rate in women over the age of 40. Many studies have shown that the detection of cancer at earlier stages significantly reduces patients' mortality and morbidity rages.
Ultrasound computer tomography (USCT) is considered as a promising screening tool for diagnosing early-stage breast cancer as it is cost-effective and produces 3D images without radiation exposure. However, USCT is not a popular choice mainly due to its prolonged imaging time. USCT is time-consuming because it needs to transmit a number of ultrasound waves and record them one by one to acquire a high-quality image.
We propose FastUSCT, a method to acquire a high-quality image faster than traditional methods for USCT. FastUSCT consists of three steps. First, it transmits multiple ultrasound waves at the same time to reduce the imaging time. Second, it separates the overlapping waves recorded by the receiving elements into each wave with UNet. Finally, it reconstructs an ultrasound image with a synthetic aperture method using the separated waves.
We evaluated FastUSCT on simulation on breast digital phantoms. We trained the UNet on simulation using natural images and transferred the model for the breast digital phantoms. The empirical result shows that FastUSCT significantly improves the quality of the image under the same imaging time to the conventional USCT method, especially when the imaging time is limited.
\end{abstract}

\section{Introduction}

Breast cancer is the most prevalent cancer with a high mortality rate in women over the age of 40 \cite{chu1996recent}. Many studies have shown that the detection of cancer at earlier stages significantly reduces patients’ mortality and morbidity rates \cite{cianfrocca2004prognostic}.
Thus, variety of breast cancer diagnosis modalities have been developed to detect cancer at early stages. 

Although conventional X-ray mammography (MG) is the most widely used cancer screening tool to date, it has several drawbacks. Because the image is a 2D projection of the breast, the lesions appear embedded in a highly heterogeneous background, which makes the detection of small lesions difficult. In addition, MG has low accuracy for younger women with dense breast tissue \cite{nothacker2009early}. Conventional hand-held ultrasound imaging performs well in dense breast tissue but its accuracy is dependent on the skill of the operator \cite{tohno2012educational}. Magnetic Resonance Imaging (MRI) can overcome these limitations but is prohibitively expensive for screening.
As such, ultrasound computer tomography (USCT) imaging is considered as a promising new imaging device for breast cancer diagnosis as it (1) produces 3D volumetric images, (2) incurs no radiation exposure, (3) is cost-efficient, (4) does not require skilled sonographers \cite{stotzka2002medical,duric2005development,gemmeke20073d,ruiter20123d,hopp2013automatic}.

\begin{figure*}
    \centering
    \includegraphics[width=0.9\textwidth]{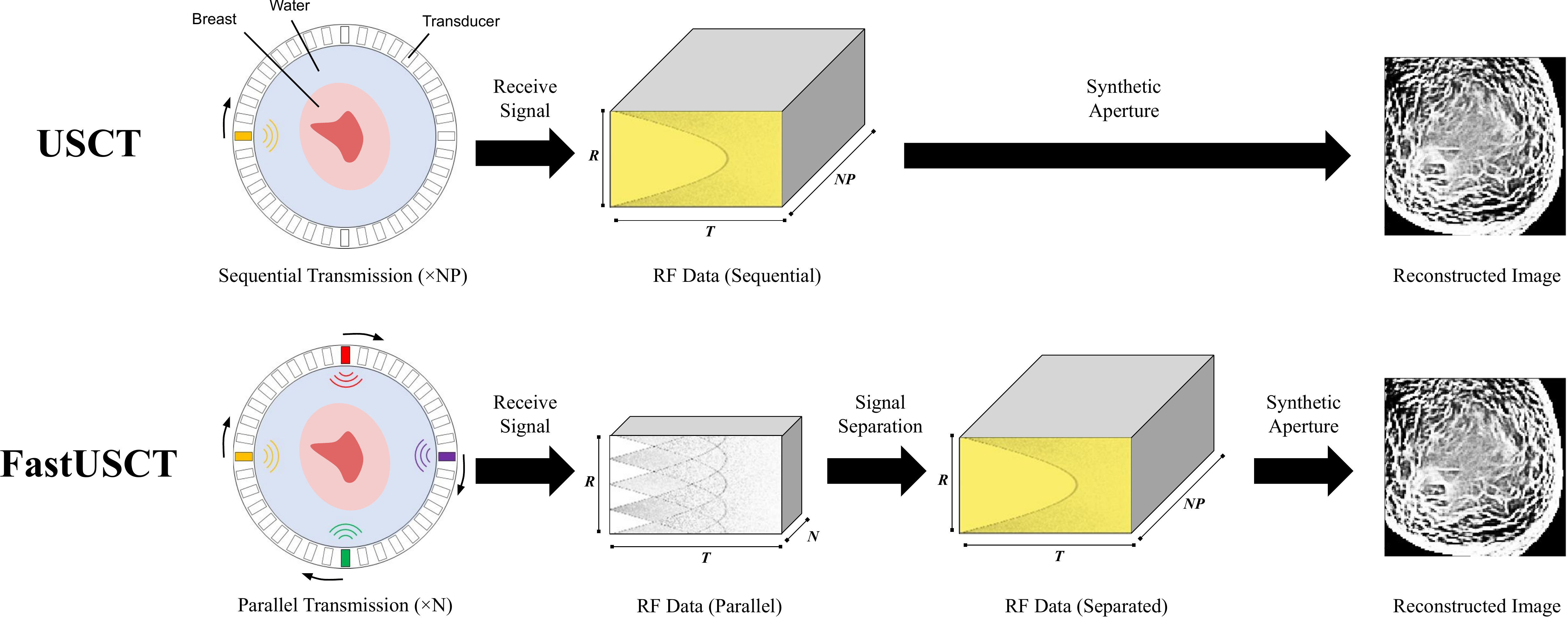}
    \caption{The overview of conventional USCT and FastUSCT. FastUSCT transmits multiple waves at once to reduce the imaging time. The overlapping waves are separated by a signal separation module to produce RF data of sequential transmission. We reconstruct the image from the separated RF data by the synthetic aperture.}
    \label{fig:fastusct}
\end{figure*}

The basic idea of USCT is to surround the subject with a number of ultrasound transducers in a water bath (Figure \ref{fig:fastusct}) \cite{greenleaf1981clinical,duric2005development,ruiter20123d}. The transducers are used to transmit and receive ultrasound waves.
Unfocussed ultrasound waves are transmitted and interact with the subject tissue. Ultrasound waves are reflected and refracted at interfaces with different acoustic properties such as density and sound speed. The interactions cause reflected as well as transmitted signals which are recorded by the ultrasound transducers from every direction. 
After that, a 2D slice image is reconstructed by synthesizing the recorded echo signals. Once a number of successive slices are collected, they can be stacked together to form a 3D image of the subject.

One of the problems of USCT is that it requires prolonged imaging time. To make a high-quality image, USCT requires a number of transmissions for each slice. Because the imaging time is roughly linear to the number of transmissions, producing a high-quality image takes prolonged time.
While MG typically takes a few seconds for the scan, USCT often takes approximately 20 minutes to scan a 3D image due to the hardware constraints \cite{gemmeke20073d}. The imaging time is crucial to the cost-effectiveness and patient acceptance of medical imaging devices. 


In this paper, we propose FastUSCT, a method to produce a high-quality ultrasound image within a shorter amount of time than USCT. 
FastUSCT simultaneously sends multiple ultrasound waves from multiple transmitting elements at once and collect the signal data (radio-frequency data; RF data) at the receiving elements to reduce the imaging time of each 2D slice. We then separate the RF data with overlapping waves into RF data for each wave.

FastUSCT has several advantages. 
First, by sending multiple ultrasound waves at once, it produces high-quality images with a short amount of time. Reducing the imaging time is especially valuable for USCT as it usually takes about 20 minutes for the imaging sequence. 
Second, because multiple signals are received at the same time, the amount of signal data required to store is reduced. Because the size of the raw signal data for one patient is roughly 1 TB, we can reduce the data size to half (500 GB) by parallelizing two waves using FastUSCT, which is already a significant improvement.
Third, because FastUSCT separates the signal data and then reconstruct images instead of directly predicting the reconstructed images \cite{wang2018image,cheng2019deep}, it can be used to generate images for several imaging modes such as brightness mode (B-mode) \cite{gemmeke20073d}, sound speed \cite{li2009vivo,nebeker2012imaging}, and wave attenuation \cite{li2008breast,li2017breast}. 
In this paper, we focus on the brightness mode as it has high spatial resolution and is the most common imaging mode for hand-held ultrasound. We use the Synthetic Aperture method (SA) \cite{jensen2006synthetic}. An image is reconstructed by synthesizing the recorded echo signals based on the delay-and-sum algorithm \cite{van1988beamforming}.
Finally, FastUSCT is not restricted to any specific subjects and applicable to any subjects the conventional USCT is applicable.


To evaluate the image quality of FastUSCT, we trained the signal separation model and tested our method on three simulation settings.
To evaluate the robustness of the method to previously unseen images, we trained the signal separation model on subjects converted from natural images and transferred the model to 50 breast digital phantoms.
The experimental results show that the proposed method significantly improves the image quality measured by structural similarity and peak-signal-to-noise compared to the sequential USCT, especially when the imaging time is short.


\section{Background}

\subsection{Ultrasound Computed Tomography (USCT)}
\label{sec:usct}

The idea of USCT is not new; one of the earliest publications of USCT is in 1981 \cite{greenleaf1981clinical}. Since then, many researchers have been developing the ultrasound tomography devices and algorithms \cite{schomberg1978improved,norton1979ultrasonic,greenleaf1982computed,ashfaq2004new,jirik2005ultrasonic,matthews2017regularized,cheng2019deep,vu2020generative}.
However, USCT is not realized for clinical practice mainly because of the time-consuming imaging time and the huge storage requirement.

The bottleneck of the imaging time of USCT is the flight time of the ultrasound waves. 
The flight time of the waves is negligible for modalities based on electromagnetic waves such as X-ray CT. However, because the speed of sound is much slower than the speed of X-ray, the flight time is rather the bottleneck of the imaging time of USCT.
As such, the imaging time of USCT is roughly linear to the number of transmission iterations.

To generate a 3D volumetric image, we repeat the 2D slice reconstruction procedure multiple times to generate successive 2D slices to generate a 3D image of the patient. In this paper, we focus on speeding up the imaging time of the 2D slice as speeding-up the reconstruction of 2D slices directly speeds up the reconstruction of a 3D image.

Although many methods have proposed to speed up the imaging time for MRI \cite{hutchinson1988fast,oshio1991grase,niendorf2006parallel,keereman2010mri,feinberg2013ultra,sun2016deep,feng2017compressed}, they are relying on the fact that MRI collects k-space data \cite{huettel2004functional} whereas USCT collects RF data. As such, fast image reconstruction methods proposed for MRI are not directly applicable to USCT.

\subsection{Signal Separation}

There are several methods to transfer multiple signals at once.
Frequency-division multiplexing \cite{weinstein1971data} transfers multiple signals at once by sending signals in several distinct frequency ranges.
The drawback of the method for ultrasound imaging is that by dividing the bandwidth for each signal, the sharpness of the pulse is degraded. As a result, the spatial resolution of the image decreases.

The other idea is to separate the overlapping signals at the receiving end by a signal separation algorithm.
Research on the signal separation is widely conducted in the fields such as speech separation and seismic signal denoising \cite{douglas2001blind,o2005survey,comon2010handbook,liu2013blind}.
Many blind signal separation methods have been proposed to separate the spectrogram of a mixed-signal based on the differences of timing and frequency \cite{o2005survey,woo2005neural,simpson2015deep,wang2018combining,wang2018supervised,comon2010handbook}, but separating two waves with the same characteristic is still difficult.

\section{Proposed Method: FastUSCT}

In this section, we first show the overview of the proposed method, FastUSCT. We then describe the detailed description of the signal separation model.

\subsection{System Overview}

FastUSCT consists of three steps.

\begin{figure*}
    \centering
    \includegraphics[width=0.8\textwidth]{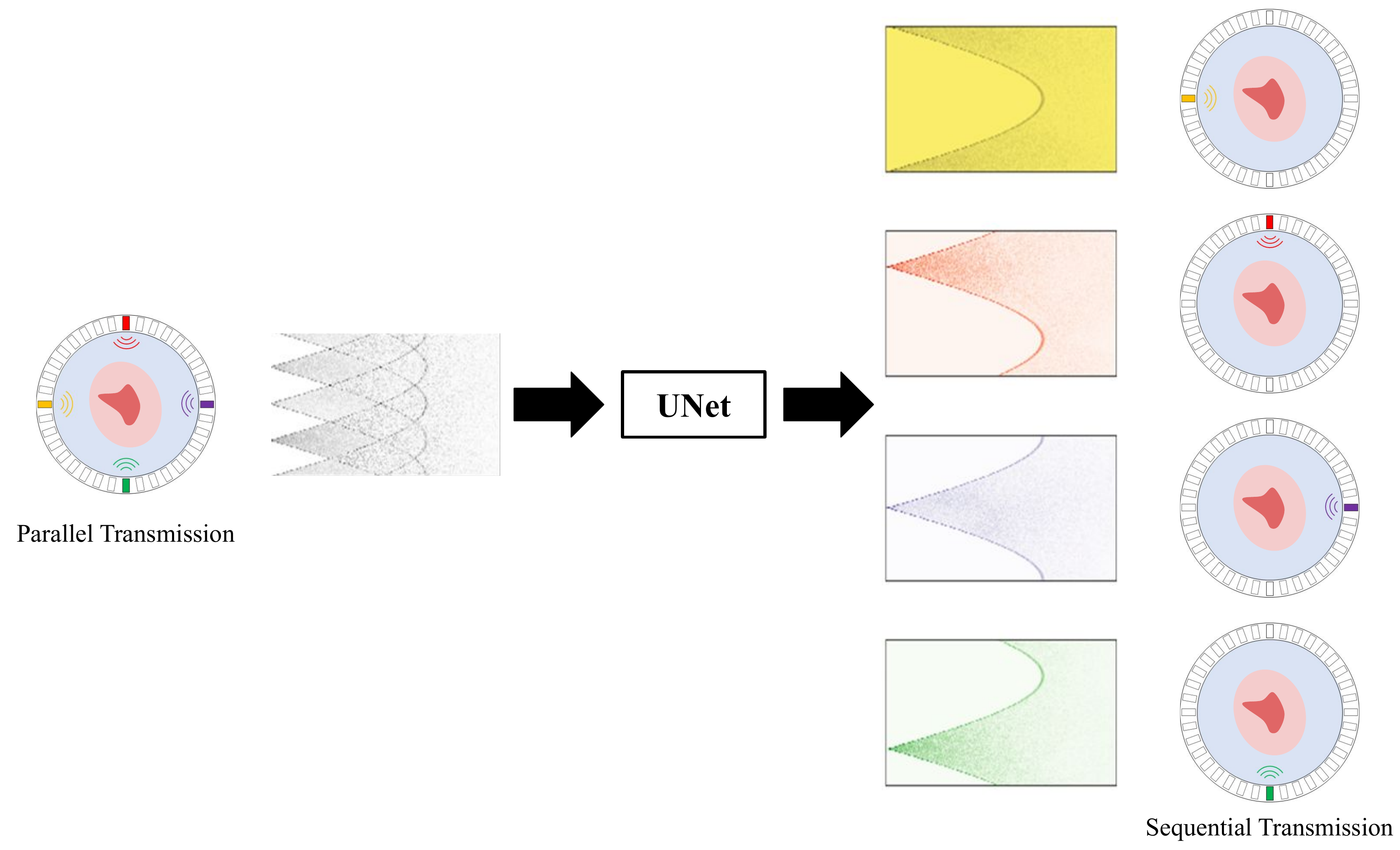}
    \caption{Signal separation model. The input is the RF data of the parallel transmission. The task is to separate the overlapping waves to each one.}
    \label{fig:signal-separation}
\end{figure*}

\begin{enumerate}
    \item {\bf Signal Collection}: First, we transmit multiple ultrasound waves at once and record the RF data, the signal data received at the receiving elements. We repeat this iteration for $N$ times to collect sufficient signals to reconstruct a high-resolution image. In total, we collect $N$ matrices of $(R, T)$-dimensions in this first step, where $N$ is the number of iterations, $R$ is the number of receiving elements, $T$ is the number of samples.
    \item {\bf Signal Separation}: Second, we separate the RF data of multiple waves into the RF data of each wave using UNet \cite{ronneberger2015u}. The signal separation model takes the RF data of parallel transmission as an input and outputs the RF data of single transmissions (Figure \ref{fig:signal-separation}). We repeat this process for the $N$ matrices collected at the first step to generate $N \cdot P$ matrices of $(R, T)$-dimensions of the separated RF data, where $P$ is the number of waves transmitted at once in each iteration.
    \item {\bf Image Reconstruction}: Finally, we reconstruct the image from the separated RF data by a SA method. 
\end{enumerate}

\subsection{Signal Separation Model}

Due to the hardware constraint of USCT system, the frequency band of the transmitted waves is the same for all the waves.
Therefore, FastUSCT has to separate overlapping waves with the same frequency bands. We use a UNet \cite{ronneberger2015u} to separate the raw RF data of the mixed signals into each one.
The input of the UNet is the raw RF data instead of the spectrogram to make use of the phase information. 
Because a slight phase error causes a huge image quality degradation in ultrasound imaging \cite{nock1989phase},
we use a loss function with emphasis on the error on the phase of the signal. The loss function is the weighted sum of mean square error (MSE) and phase MSE (PMSE):

\begin{align}
    Loss &= MSE + \alpha PMSE \nonumber \\
         &= \frac{1}{N} \sum_{i=1}^{N} (y_i - \hat{y_i})^2 \nonumber \\
            + \alpha & \cdot \frac{1}{N} \sum_{i=1}^{N} ((cos \theta_i - cos \hat{\theta_i})^2 + (sin \theta_i - sin \hat{\theta_i})^2 ),
\end{align}

where $N$ is the number of signals, $\alpha$ is the weight of the PMSE, $y_i$ and $\hat{y_i}$ are the prediction (separated RF data) and the label (RF data of sequential transmission) of $i$-th datapoint, respectively. $\theta_i$ and $\hat{\theta_i}$ are the phases of $y_i$ and $\hat{y_i}$, respectively. The phases of signals are calculated as follows:

\begin{equation}
    \theta_i = tan^{-1} \frac{\bar{y_i}}{y_i},
\end{equation}

where $y_i$ and $\bar{y_i}$ are the original signal and its Hilbert transformation \cite{hahn1996hilbert}, respectively, $\theta_i$ is the calculated phase of $y_i$.

\section{Experiments}

We ran experiments on four simulations to evaluate the performance of FastUSCT.
We first ran simulations on two settings with scatterers: scatterer points simulation (Section \ref{sec:scatterer-points}) and scatterer clump simulation (Section \ref{sec:scatterer-clump}). We randomly generated simple subjects to quantitatively evaluate the accuracy of the signal separation module in simple controlled experiments.
Next, we ran a simulation using density maps converted from natural images to evaluate the performance of FastUSCT in a broad range of subjects (Section \ref{sec:natural-image}).
Finally, we evaluated the performance of the FastUSCT on breast digital phantoms (Section \ref{sec:digital-phantom}). We transferred the signal separation model trained on the natural images to the breast digital phantoms to evaluate the robustness of the proposed method.

We denote FastUSCT with $N$ iterations of transmissions each consists of $P$ waves by FastUSCT($N$, $P$) (thus a total of $N \cdot P$ waves), and USCT with $N$ iterations of sequential transmissions by USCT($N$).

\subsection{Simulation Setup}
\label{sec:simulation}

\begin{table}
    \centering
    \begin{adjustbox}{width=\columnwidth}
    \begin{tabular}{c|c}\toprule
    Parameter [unit] (notation) & Value \\ \midrule
    Radius of ring array transducer [mm] & 	50 \\
    Number of receiving elements ($R$) & 256 \\
    Sampling frequency [MHz]	& 40 \\
    Number of samples ($T$)	& 4096 \\
    Sound speed [m/s]	& 1450 \\
    Average density [kg/m$^3$] ($d_0$)	& 1000 \\ 
    Pixel size [mm]	& 0.125 \\
    Pixel resolution of simulation grid	& 1024 $\times$ 1024 \\
    Center frequency of transmitted waves [MHz]	& 2 \\
    Number of cycles of transmitted waves	& 1 \\
    \bottomrule
    \end{tabular}
    \end{adjustbox}
    \caption{Parameters of the simulation on k-Wave toolbox \cite{treeby2010k}.}
    \label{tab:usct-parameters}
\end{table}

Given the density map of the subject, we ran a wave simulation with the parameters listed in Table \ref{tab:usct-parameters}.
RF data was acquired by a SA method that collects the RF signals for all the combinations of the transmission iterations and the receiving elements. 
Therefore, acquired RF data has three dimensions, transmission iteration $N$, receiving elements $R$, and samples $T$.

We ran the simulation using a USCT simulator \cite{tomii} based on the k-Wave toolbox \cite{treeby2010k}. k-Wave is a simulation tool widely used for USCT \cite{matthews2017regularized,cheng2019deep,vu2020generative} as it is an open source software for the time-domain simulation of propagating acoustic waves with a sophisticated numerical model that can account for both linear and nonlinear wave propagation \cite{treeby2010modeling,treeby2012modeling}.


\subsection{Scatterer Points Simulation}
\label{sec:scatterer-points}

\begin{figure*}[hbt]
    \centering
    \includegraphics[width=0.95\textwidth]{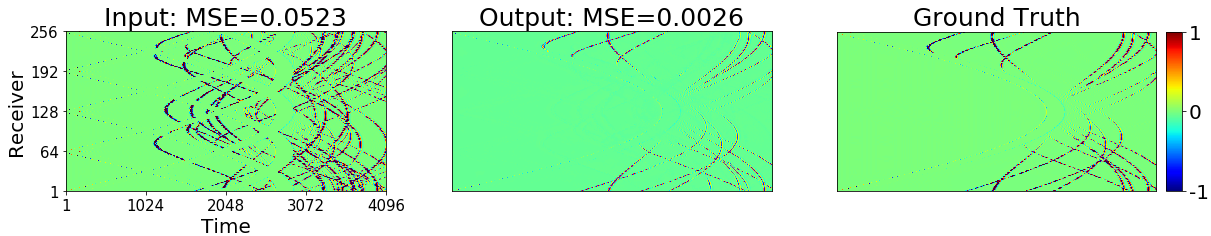}
    \caption{RF data separated using UNet on scatterer points simulation. The input is the RF data collected by FastUSCT(4, 4). The output is the RF data after the signal separation. The ground truth is the label of the signal separation (i.e. RF data of USCT(16)). The horizontal axis is the sample axis $T$ and the vertical axis is the receiving elements $R$.}
    \label{fig:scatterer-signal}
\end{figure*}

To evaluate the proposed method in a controlled environment, we ran simulations with randomly placed scatterer points. 
We ran 100 simulations on scatterer subjects to train the signal separation model. 
For each simulation, we picked 10 positions uniformly randomly from $(-256, 256) \times (-256, 256)$ grid where $x=0, y=0$ being the center of the simulation grid.
We placed a scatterer with a density of $1500 kg/m^3$ at the picked positions. A scatterer is an object with a high density thus it generates scattered waves when it is hit by a wave.

We ran simulations of sequential transmission and parallel transmission on each generated subject and collected their RF data.
We set the number of simultaneously transmitting elements to 4 for the parallel transmission simulations. Because the total number of transmitting elements is 16, four sets of transmissions are done.
Each data was augmented based on the symmetry of the ring array elements. Because the elements were arranged in reflection symmetry, the data was augmented by reversing the order of the elements. Furthermore, because the transmitting elements array was in four-fold rotational symmetry, the data was augmented by shifting the order of the elements for three times for every 90 degrees. 
Therefore, $4 \times 2 \times 4$ datapoints for the signal separation model were generated from a single simulation.
Thus, the dataset consists of a total of 100 $\times 32$ RF data collected from 100 simulations. 
The ground truth label of the signal separation is the RF data collected by a simulation of sequential USCT.
We divided the dataset into a training dataset consists of 2560 datapoints from 80 simulations and a validation dataset consists of 640 datapoints from 20 simulations.

Because raw RF data is noisy and has high variance, we preprocessed the signal data to normalize the output range to $(-1, 1)$. See the supplementary material for the detailed desciription.
We trained the model using Adam optimizer \cite{kingma2014adam} with the coefficients $\beta_1=0.9$ and  $\beta_2=0.999$. 
We set the weight $\alpha$ of the phase MSE to $0.01$. 
The initial learning rate was set to $0.01$, multiplied by a factor of 0.9 every 10 epochs. We trained the model for 100 epochs. We set the batch size to 8. 

We evaluated the performance of FastUSCT on 50 scatterer subjects randomly generated. We evaluated the accuracy of the separated RF data and the quality of the images reconstructed from the separated RF data. The mean squared error of the separated RF signal from the corresponding label was $3.086 \times 10^{-3}$ on average with $SD = 0.506 \times 10^{-3}$.

We reconstructed the ultrasound images of the patch of size $512 \times 512$ pixels at the center of the simulation grid from the separated signals using delay-and-sum beamforming \cite{van1988beamforming}. The resolution of the reconstructed image was $256 \times 256$.

We followed the conventional postprocessing for ultrasound images and applied a log compression \cite{seabra2008modeling}, cut-off to -30 dB to 0 dB, and then applied 3 $\times$ 3 averaging filter.
We evaluated the quality of the images by mean structural-similarity (MSSIM) and peak-signal-to-noise-ratio (PSNR) compared to the image reconstructed by USCT(16) \cite{wang2004image}.
We set the window size of MSSIM to $7 \times 7$.

\begin{table*}[hbt]
    \centering
    \begin{tabular}{c|cc} \toprule
        Scatterer Points     & Without separation & With separation \\ \midrule
        RF MSE     & $5.313 \times 10^{-2} (0.241 \times 10^{-2}) $  & $\mathbf{0.309 \times 10^{-2} (0.005 \times 10^{-2})}$ \\
        Image MSSIM & $0.840 (0.080)$ & $\mathbf{0.998 (0.002)}$ \\ 
        Image PSNR & $33.54 (2.16)$ & $\mathbf{47.30 (2.38)}$ \\
        \bottomrule
        \toprule
        Scatterer Clump     & Without separation & With separation \\ \midrule
        RF MSE     & $6.285 \times 10^{-2} (0.071 \times 10^{-2}) $ & $\mathbf{1.908 \times 10^{-2} (0.074 \times 10^{-2})} $ \\
        Image MSSIM & $0.313 (0.035)$ & $\mathbf{0.590 (0.124)}$ \\ 
        Image PSNR & $19.60 (0.45)$ & $\mathbf{23.76 (1.28)}$ \\ \bottomrule
    \end{tabular}
    \caption{The performance of FastUSCT(4, 4) with and without signal separation on scatterer points simulation and scatterer clump simulation compared to USCT(16). The values are the average over the simulations with the standard deviation in the parentheses. RF MSE: The mean squared error of the separated RF data to the RF data of the sequential transmission. Image MSSIM: MSSIM of the reconstructed image. The window size of MSSIM is $7 \times 7$. Image PSNR: PSNR of the reconstructed image.}
    \label{tab:scatterer-image}
\end{table*}

Table \ref{tab:scatterer-image} summarizes the image quality of the reconstructed images by FastUSCT. Both MSSIM and PSNR were significantly improved by the signal separation.
Figure \ref{fig:scatterer-signal} is an example of the RF data recorded by FastUSCT(4, 4). The horizontal axis is the samples $T$ and the vertical axis is the receiving elements $R$.
The left figure is the RF data of FastUSCT(4, 4) corresponding to the first transmission wave. The middle figure is the separated RF data of FastUSCT(4, 4) corresponding to the first transmission wave. The right figure is the RF data of USCT(16) corresponding to the first transmission wave. 
The result shows that the model successfully separated the mixed signals into each one. The reconstructed images are in the supplementary materials.

\subsection{Scatterer Clump Simulation}
\label{sec:scatterer-clump}

Next, we evaluated the performance of the proposed method on a clump of scatterers. The signal separation task is more complicated for a clump of scatterers because the scattering waves from the scatterers are more intertwined in this setting.

The clump of scatterers was generated as follows.
First, we pick an intensity uniformly random from $(0, 1)$ for each $16 \times 16$ patch of pixels in the simulation grid. Then, the intensity of the pixels outside a circle with a center position at (-64, 0) with a radius of 64 pixels were set to 0.
Finally, the intensity of the pixels were converted into the density map as:

\begin{equation}
\label{image2density}
    D(i, j) = d_0 + a \cdot \epsilon_{i, j} \cdot I(i, j)
\end{equation}

where $D(i,j)$ is the density value of the $(i, j)$-th pixel, $d_0$ is the average density, $a$ is the coefficient that determines the density variation. We set $a$ to 100 to make the density range similar to the actual breast \cite{duck2013physical}. See Table \ref{tab:usct-parameters} for the other simulation parameters.
$\epsilon_{i, j}$ is a random value sampled from a normal distribution $\mathcal{N}(0, 1)$, and $I(i,j)$ is the intensity of the $(i, j)$-th pixel. 
The other procedures on the training and the evaluation were the same as the scatterer points simulation.

\begin{figure*}
    \centering
    \includegraphics[width=0.75\textwidth]{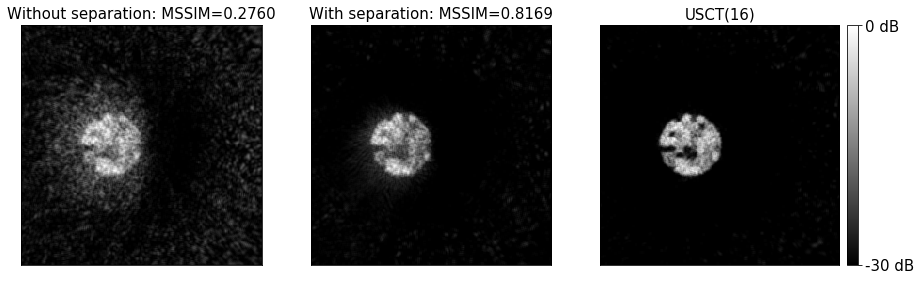}
    \caption{Image reconstructed by FastUSCT(4, 4) compared to the USCT(16) which requires four times of imaging time on scatterer clump simulation. Left: reconstructed from RF data without separation. Middle: reconstructed from the separated RF data. Right: reconstructed by USCT(16).
    }
    \label{fig:clump-image}
\end{figure*}

Table \ref{tab:scatterer-image} summarizes the results. Even with a clump of scatterers, the RF data was successfully separated and the RF MSE was significantly reduced. See the supplementary material for the visual comparison of the RF data.
Figure \ref{fig:clump-image} is an example of the reconstructed images with and without the signal separation. The quality of the image reconstructed with the separation was close to the image reconstructed by USCT(16) which takes four times of imaging time.

\subsection{Natural Image Simulation}
\label{sec:natural-image}

To evaluate the performance of the proposed method on a wide variety of subjects, we tested its performance on the subjects converted from the images retrieved from the ImageNet \cite{deng2009imagenet}.

We converted the images into density maps as follows. First, the image was resized to the same size as the simulation grid (1024 $\times$ 1024), converted to grayscale, and normalized in the range from 0 to 1. Then, the intensity of the pixels outside the ring array were set to 0. Finally, the image was converted into the density map by (Equation \ref{image2density}).

\begin{figure}[htb]
    \centering
    \subfloat[ImageNet]{\includegraphics[width=0.9\columnwidth]{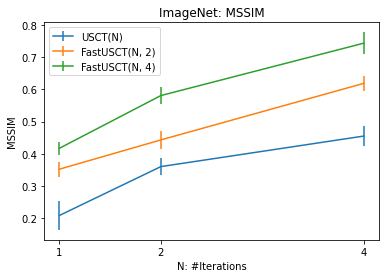} \label{fig:imagenet-ssim}}

    \subfloat[Breast Digital Phantom]{\includegraphics[width=0.9\columnwidth]{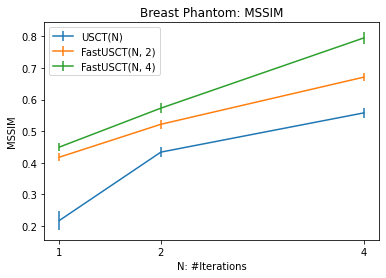} \label{fig:breast-ssim}}

    \caption{Comparison of the image quality (MSSIM) on natural image simulation and breast digital phantoms. The error bar shows the standard deviation.}
    \label{fig:ssim}
\end{figure}
\begin{figure*}[hbt]
    \centering
    \includegraphics[width=0.7\textwidth]{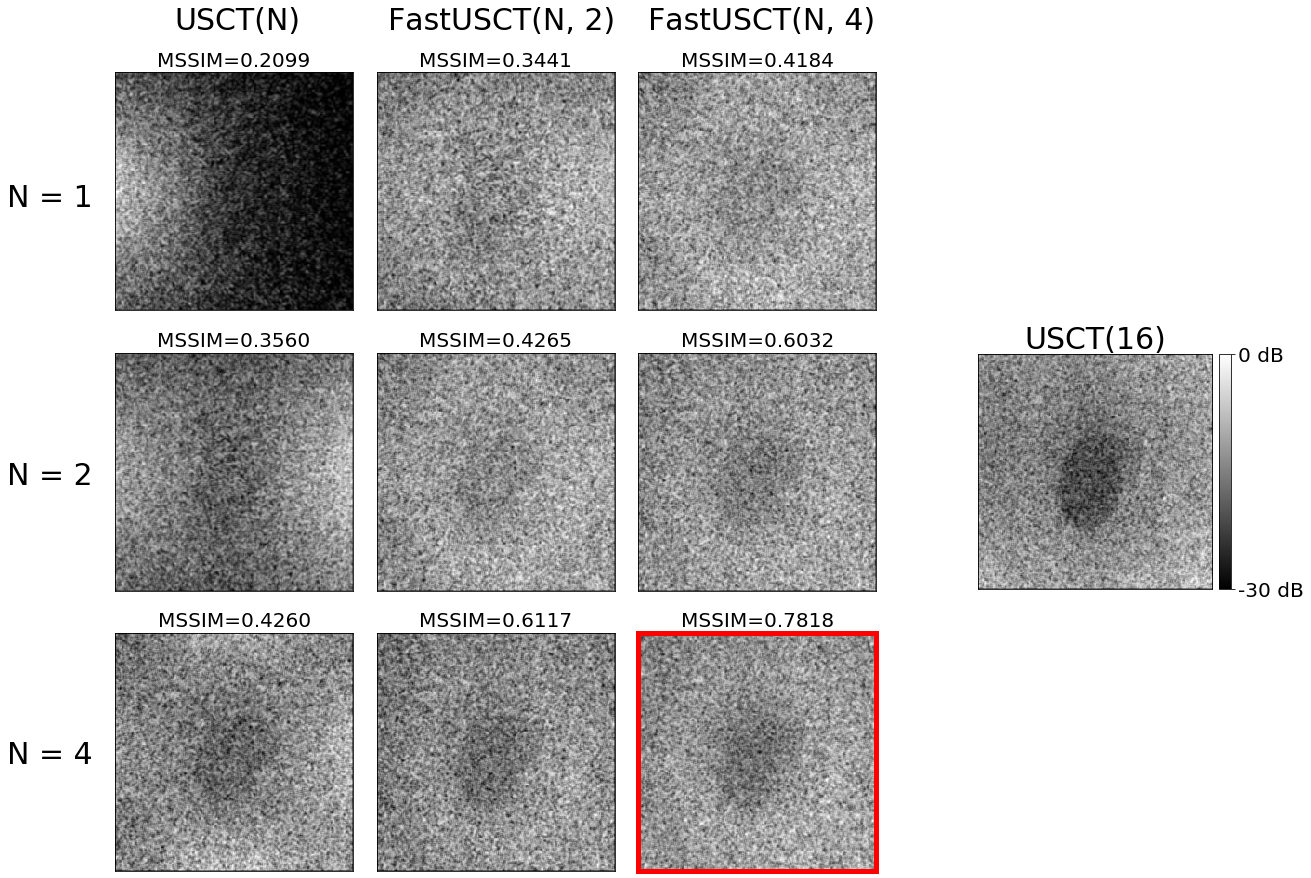}
    \caption{Example of images reconstructed by FastUSCT on natural image simulation. FastUSCT(4, 4) had the highest MSSIM.}
    \label{fig:imagenet}
\end{figure*}

We trained the model on 30400 datapoints from 950 simulations and evaluated on 50 simulations. 
We ran experiments with $N = 1, 2, 4$ and $P = 1, 2, 4$.
The MSE of the RF data to the label was $1.480 \times 10^{-3}$ on average with standard deviation $SD = 0.839 \times 10^{-3}$ for FastUSCT(4, 2) and $1.918 \times 10^{-3}$ on average with $SD = 1.136 \times 10^{-3}$ for FastUSCT(4, 4).
Figure \ref{fig:imagenet-ssim} compares MSSIM of FastUSCT to the image reconstructed by USCT(16). The comparison of PSNR is in the supplementary material. The result shows that under the same imaging time, MSSIM and PSNR are significantly higher on FastUSCT compared to USCT with the same imaging time. 
Also note that the MSSIM of FastUSCT(1, 4) is higher than that of USCT(2), and on par with that of USCT(4). The result shows that FastUSCT has successfully speeded-up the imaging time by a factor of four with a marginal drop of MSSIM.
The improvement of the visual quality is especially clear when the number of transmission iteration is 1 (Figure \ref{fig:imagenet}).

\subsection{Transfer to Breast Digital Phantoms}
\label{sec:digital-phantom}

Finally, we evaluated the performance of the proposed method on 50 breast digital phantoms. See the supplementary material for the details of the phantoms.
We used the signal separation model trained in Section \ref{sec:natural-image} and transferred it to breast digital phantoms.
Medical images tend to be too small in sample size to train a neural network. Our goal here is to evaluate the possibility of using natural images to improve the quality of breast images.

\begin{figure*}
    \centering
    \includegraphics[width=0.7\textwidth]{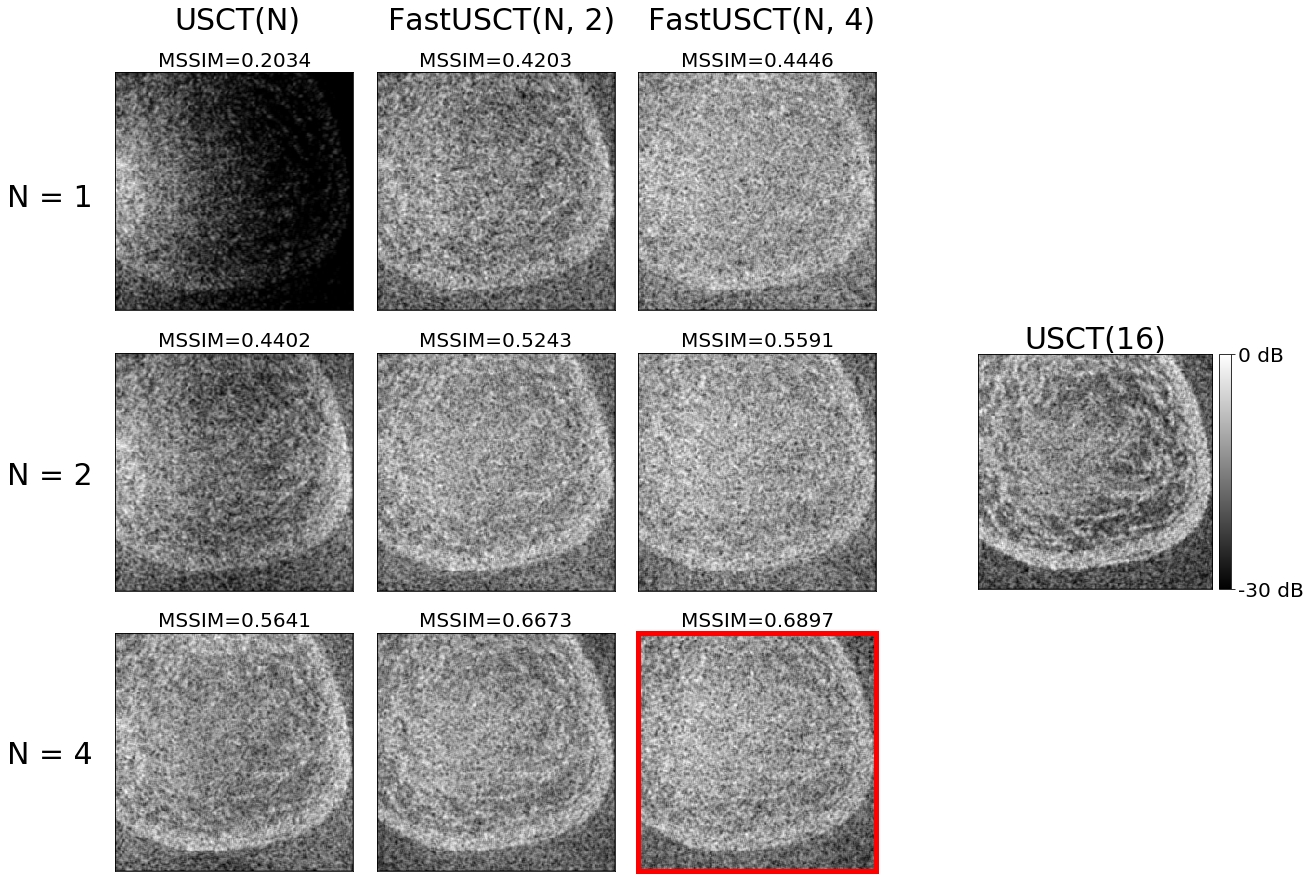}
    \caption{Example of breast phantom images reconstructed by FastUSCT. FastUSCT(4, 4) had the highest MSSIM.}
    \label{fig:phantom}
\end{figure*}

We ran simulations of FastUSCT with $N = 1, 2, 4$ and $P = 1, 2, 4$ to reconstruct images of 50 breast digital phantoms.
The MSE of the RF data to the label was $5.958 \times 10^{-3}$ on average with $SD = 0.850 \times 10^{-3}$ for FastUSCT(4, 2) and $7.254 \times 10^{-3}$ on average with $SD = 1.085 \times 10^{-3}$ for FastUSCT(4, 4).
Figure \ref{fig:breast-ssim} shows the comparison of MSSIM (See the supplementary material for PSNR). FastUSCT achieved significantly higher image quality compared to USCT with the same imaging time. Moveover, MSSIM of FastUSCT(1, 4) was higher than that of USCT(2). The result shows that FastUSCT has successfully speeded-up the imaging time by a factor of two without a drop in MSSIM even when the training data is substantially different from the test dataset.
Figure \ref{fig:phantom} is an example of the reconstructed image of the breast digital phantom. 

\section{Conclusions}

In this paper, we proposed FastUSCT, a method to improve the imaging time of USCT imaging by firing multiple transmitters and separate the acquired signals by UNet. We ran four experiments on simulation and showed that FastUSCT improves the image quality compared to USCT with the same imaging time.
We trained the signal separation model on natural images and then transferred the model to breast digital phantoms. The empirical result on breast digital phantoms showed that FastUSCT significantly improves the image quality compared to USCT with the same imaging time.

\clearpage


\bibliography{aaai21}
\bibliographystyle{icml2021}

\clearpage
\appendix

\section{Preprocessing the Signal}

We describe the preprocess applied to the RF data to normalize the data.
We first applied a band-pass filter with a cutoff frequency of 0.5 MHz and 8.0 MHz. 
Then we masked the parts of RF data which contains non-scattering waves to 0. Because the amplitude of the non-scattering waves is much larger than the scattering waves, it significantly degrades the quality of the separation. Because non-scattering waves are just bypassing the subject, they have no information about the position of the subject. Thus, masking out these waves have no negative effect on the reconstructed images. Masking was performed by the following two processes. First, the envelope of the signal was detected using Hilbert transformation:

\begin{equation}
    e_i = \sqrt{x_i^2 + \bar{x_i}^2}
\end{equation}

where $x_i$ and $\bar{x_i}$ are the original signal and its Hilbert transformation, respectively.
Then, the point with maximum amplitude was found, and the signals within a certain range from the maximum point were masked to 0.
For the scatterer points simulation and scatterer clump simulation, we set the range of the mask from 100 time-steps before to 120 time-steps after the maximum point. 
For the natural image simulation and breast digital phantom, we set the range of the mask from 30 time-steps before to 50 time-steps after the maximum point.
Finally, to reduce the variance of the RF data, we divided the amplitude of each RF data by 3 times the standard deviation of each RF data and cut-off the value to $(-1, 1)$.

\section{Hyperparemeter of the Experiment}

We use a UNet \cite{ronneberger2015u} to separate the raw RF data of the mixed signals into each one.
UNet consists of convolution layer, max pooling layer, up-convolution layer, and skip connection. For all the convolution layers, zero-padding was applied to the time direction, and circular-padding was applied to the receiver direction because the first element and the last element are next to each other in the ring array. Dilated convolution \cite{yu2015multi} with the dilation rate of 1 and 4 in receiver direction and time direction, respectively, was applied for all convolution layers because the sampling interval is short compared to the receiver pitch, and the same wave appears at a distant position in the time direction. All the convolution layers except the output layer were followed by Batch Normalization layer \cite{ioffe2015batch} and ReLU activation \cite{nair2010rectified}. The convolution layer in the output layer was followed by Tanh activation to normalize the output to $(-1, 1)$. 

We implemented the signal separation model using the PyTorch library \cite{paszke2019pytorch}. We ran the experiments on Amazon EC2 P3 instance with NVIDIA V100 GPUs with 32 GB GPU memory.

\begin{figure}
    \centering
    \includegraphics[width=0.96\columnwidth]{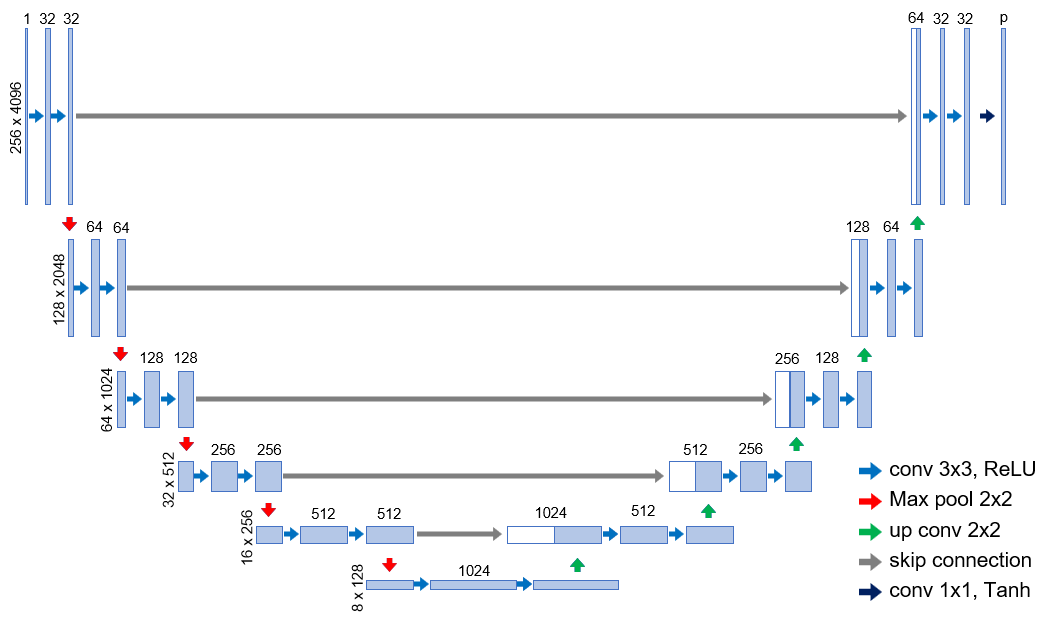}
    \caption{UNet architecture used for the signal separation.}
    \label{fig:unet}
\end{figure}





\section{Breast Digital Phantom}

We generated breast digital phantoms from real breast ultrasound images. We processed the ultrasound images to generate the density map as follows. First, we resized the image to the simulation grid size ($1024 \times 1024$).
Then, we segmented the breast region from the image by snake active contour to remove the ultrasound artefacts surrounding the breast region \cite{kass1988snakes}. We sharpened the image twice and then applied a contrast enhancement by S-curve \cite{ahmed2011improving,gandhamal2017local}. Finally, we converted the resulting image into a density map by Equation 3 in the main paper.

\noindent{\bf Sharpening}:
\begin{enumerate}
    \item Apply a 5x5 mean filter to the input image to generate a blurred image.
    \item 
    Generate a sharpened image by following: $I_S(i, j) = 2 I_I(i, j) - I_B(i, j)$, where $I_S$ is the intensity of the sharpened image, $I_I$ is that of the input image, and $I_B$ is that of the blurred image.
    \item Cutoff the sharpened image to $(0, 1)$.
\end{enumerate}

\noindent{\bf S-curve contrast enhancement}:
\begin{enumerate}
    \item Transform the intensity of each pixel by $1 / (1 + \exp{(a (x - I(i, j)))})$, where $a=20, x=0.5$.
    \item Normalize the intensity linearly to $(0, 1)$.
\end{enumerate}

\section{Experimental Results}

\begin{figure}
    \centering
    \subfloat[ImageNet]{\includegraphics[width=0.95\columnwidth]{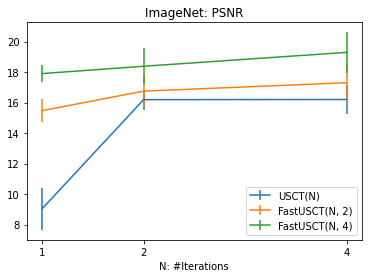} \label{fig:imagenet-psnr}}
    
    \subfloat[Breast Phantom]{\includegraphics[width=0.95\columnwidth]{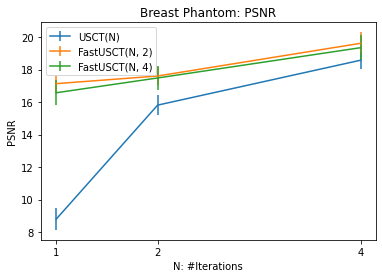} \label{fig:breast-psnr}}
    \caption{Comparison of PSNR on ImageNet and breast digital phantom simulation.}
    \label{fig:psnr}
\end{figure}

\begin{figure*}
    \centering
    \includegraphics[width=0.95\textwidth]{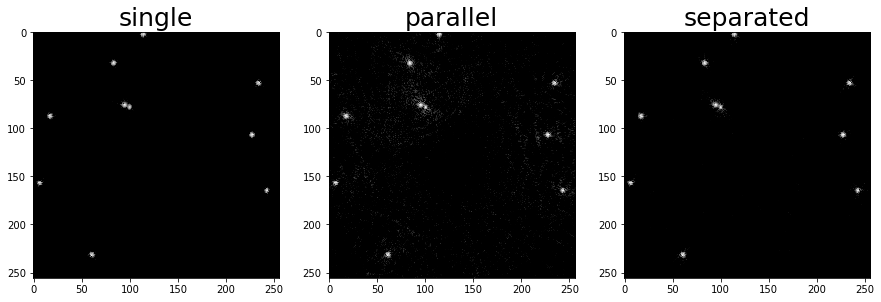}
    \caption{Example of images reconstructed by FastUSCT(4, 4) and USCT(16) on scatterer points simulation. The image quality of FastUSCT(4, 4) is close to USCT(16) which requires four times of imaging time.}
    \label{fig:scatterer-image}
\end{figure*}

\begin{figure*}
    \centering
    \includegraphics[width=0.95\textwidth]{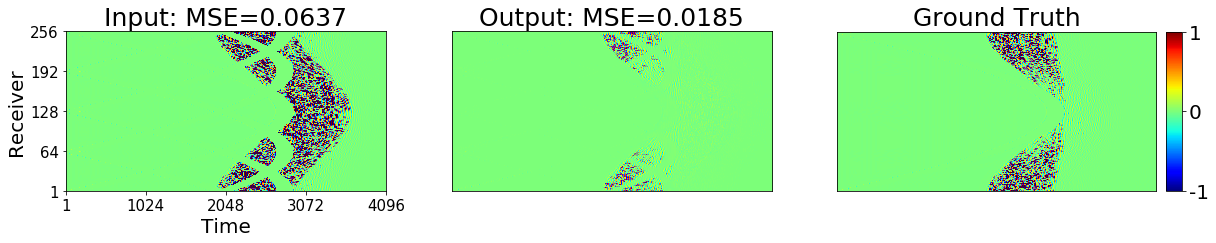}
    \caption{RF data separated using UNet on scatterer clump simulation. The input is the RF data collected by FastUSCT(4, 4). The output is the RF data after the signal separation. The ground truth is the label of the signal separation (i.e. RF data of USCT(16)). The horizontal axis is the time $T$ and the vertical axis is the receiving elements $R$.}
    \label{fig:clump-signal}
\end{figure*}

Figure \ref{fig:scatterer-image} is the comparison of the images reconstructed by FastUSCT(4, 4) and USCT(16) on scatterer points simulation.
Figure \ref{fig:clump-signal} shows the comparison of RF data on scatterer clump simulation. The result shows that the RF data is 
Figure \ref{fig:imagenet-psnr} and \ref{fig:breast-psnr} are the comparison of PSNR on ImageNet and breast digital phantoms.

\end{document}